\documentstyle[12pt]{article}
\def\picture #1 by #2 (#3){
  \vbox to #2{
  \hrule width #1 height 0pt depth 0pt
  \vfill
  \special{picture #3}}}
\def\scaledpicture #1 by #2 (#3 scaled #4){{
\dimen0=#1 \dimen1=#2
\divide\dimen0 by 1000 \multiply\dimen0 by #4
\divide\dimen1 by 1000 \multiply\dimen1 by #4
\picture \dimen0 by \dimen1 (#3 scaled #4)}}
\author{Serge Galam and Alain Mauger\\
Laboratoire des Milieux D\'{e}sordonn\'{e}s et H\'{e}t\'{e}rog\`{e}nes
\footnotemark[1]\\
Tour 13 - Case 86, 4 place Jussieu, \\ 75252 Paris Cedex 05, France\\[1ex].}
\title{A Universal Formula for~Percolation Thresholds\\ II. Extension
to Anisotropic and Aperiodic Lattices  }
\addtocounter{footnote}{0}
\footnotetext{Laboratoire de l'Universit\'{e}
P. et M. Curie - Paris 6, associ\'{e} au CNRS (URA n$^{\circ}$ 800)}
\addtocounter{footnote}{1}
\date{Phys. Rev. E, 1997}
\begin{document}
\maketitle
\baselineskip 3.3ex
\footskip 5ex
\parindent 2.5em
\abovedisplayskip 5ex
\belowdisplayskip 5ex
\abovedisplayshortskip 3ex
\belowdisplayshortskip 5ex
\textfloatsep 7ex
\intextsep 7ex
\begin{center}
{\em PA Classification Numbers:\/} 64.60 A, 64.60 C, 64.70 P\\
\end{center}

%%%%%%%%%%%%%%%%%%%%%%%%%%%%%%%%%%%%%%%%%%%%%%%%%%%%%%%%%%%%%%%%%%%%%%%
\begin{abstract}

In a recent paper, we have reported a universal power law for both site
and bond percolation thresholds in
any Bravais lattice with $q$ equivalent nearest neighbors in dimension $d$.
We now extend it to three different classes of lattices which are,
respectively,
anisotropic lattices whith not equivalent
nearest neighbors,
non-Bravais lattices with two atom unit cells, and quasicrystals.
The investigation is focussed on $d=2$ and $d=3$, due to the lack of
experimental data at higher dimensions. The extension to these lattices
requires
the substitution of $q$ by an effective (non integer) value $q_{eff}$ in the
universal law. For each out of 17 lattices which constitute our sample, we
argue
the existence of one $q_{eff}$ which reproduces both the site and the
percolation threshold, with a deviation with respect to numerical
estimates which does not exceed $\mp 0.01$.
\end{abstract}
%%%%%%%%%%%%%%%%%%%%%%%%%%%%%%%%%%%%%%%%%%%%%%
\newpage

\section{Introduction}

Very recently we have posulated a universal power law for both site and
bond percolation
thresholds [1]. The formula yields thresholds for any Bravais lattice, at
any dimension
with an impressive accuracy. It writes
\begin{equation}
p_c=p_0[(d-1)(q-1)]^{-a}d^{\ b}
\end{equation}
with $d$ the space
dimension and $q$ the coordination number.
While $b=a$ for bond dilution, it is $b=0$ for site dilution.
Three different classes were found with three distinct parameter sets
$\{p_0; \ a\}$.
The first class includes two-dimensional triangle, square
and honeycomb lattices. It is characterized by
$\{p_0=0.8889; \ a=0.3601\}$ for
site dilution and by
$\{p_0=0.6558; \ a=0.6897\}$ for bond dilution.
Two-dimensional Kagom\'{e}
and all
 other lattices of cubic symetry (for
$3\leq d\leq 6$) constitute the second class which is
 characterized by
$\{p_0=1.2868; \ a=0.6160\}$ and
$\{p_0=0.7541; \ a=0.9346\}$ for sites and bonds respectively.
A third class has been found at high dimensions ($d>6$),
which recovers the infinite Cayley tree limit, but is not relevant to
present investigation which deals with lattices only at $d=2$ and $d=3$.

Besides the dimension $d$, percolation thresholds within a class depend
only on $q$. This is understood in lattices where the $q$ nearest neighbors
of any site are equivalent, which is indeed the case for all lattices above
mentioned. This is however a drastic restriction since many percolation
problems in physics deal with lattices which do not have this property.
It is the purpose of the present work to investigate the extension of Eq. (1)
to other lattices, via the substitution of $q$ by an
effective parameter $q_{eff}$.

We have checked on a sample of 17 lattices that,
for each of them, there exists one value of $q_{eff}$ which reproduces both
the
site and the bond percolation thresholds. The error is within $\pm 0.006$ for
all the lattices,
except for two of them (dual of Penrose, and dodecagonal lattice with
ferromagnetic links), where the error reaches 0.01 for reasons
we also discuss in this paper. We consider this sample is large enough to be
representative of a general trend, and then conclude in the existence of such
a $q_{eff}$ in any lattice. For
a lattice which does not belong to our sample, this parameter could be used as
an intermediate quantity to predict bond from site percolation thresholds, or
vice versa, with the same accuracy.

The paper is organised as follows. Section 2 discusses a series of lattices
with
non-equivalent neigbhors. The introduction of an effective number of
nearest neighbors
in our universal formula for percolation is argued in Section 3. The
results are shown
and discussed in Section 3. The last section contains some remarks for
future work.

%%%%%%%%%%%%%%%%%%%%%%%%
\section{Non-equivalent neigbhors}

Some periodic Bravais lattices are anisotropic, for instance
the hexagonal lattice at $d=3$.
In this case, any lattice site has $6$ equivalent
nearest neighbors in the $a,\ b$ plane (bonding angle is $60^0$) and $2$
non-equivalent sites along the $c$ axis (bonding angle is $90^0$).
Actually, the anisotropic lattice percolation threshold should depend on
the degree
of anisotropy. For the hexagonal lattice, it means $p_{c}$ should be
different from that of the corresponding isotropic lattice with the same set
${d=3,q=8}$, i.e., the $bcc$ lattice. This difference has been observed indeed
recently [2] in the particular case of the stacked triangular lattice, which
becomes the hexagonal lattice when $a=b=c$.

There also exist non-Bravais lattices which are periodic like $fcc$ and $bcc$.
Another case is the  hexagonal close packed ($hcp$) lattice which,
on a topologic view point, is a simple hexagonal lattice
with two atoms per unit cell. Percolation thresholds
for the $hcp$ were obtained long ago [3].

Some lattices are not even periodic. This is the case of
quasicrystals which are aperiodic lattices with long-range order. Besides their
own interest such
structures can serve as models for alloy-like materials making
 a growing interest in the
determination of the quasicrystalline lattice percolation thresholds.
First determinations of percolation thresholds on $2$-$d$ quasilattices have
been made on the Penrose tiling and its dual [4, 5, 6]. Recently, percolation
thresholds have also been computed in two of the most important quasilattices,
the simple octagonal and dodecagonal tilings [7], which belong to the Penrose
local isomorphism class.

By analogy with other lattices, one can require that
lattice sites are connected only via the tile edges. The corresponding
percolation problem is named chemical percolation. Adding connection through
the diagonals of the tiles which are shorter than the tile edge leads to the
so-called ferromagnetic percolation [8]. Both these percolations will be
considered here. These lattices are reproduced in Fig. (1).

At last, some lattices do not have a single-valued coordination number q.
Lattices
with mixed-valued coordination can be either periodic or non-periodic. An
exemple
of periodic lattice at $d=2$ is provided by the dice lattice which mixes
$q=3$ and $6$. An exemple of non-periodic lattice is the Penrose
tiling, which mixes $q=3,\ 4,\ 5,\ 6$ and $7$. Note the dual of dice is the
Kagom\'e
lattice and vice versa, while the dual of the Penrose tiling is named dual
lattice of Penrose. Both the Kagom\'e and the dual of Penrose
are lattices with the single-valued coordination $q=4$. Octagonal and
dodecagonal lattices have mixed-value coordinations. Numerical estimates of
the percolation thresholds of dice can be found in [6].

%%%%%%%%%%%%%%%%%%%%%%%%%%%%%
\section{An effective number of nearest neigbhors}
In the past, attempts have been made to generalize empirical relations such
as the
Scher and Zallen approximate [9] which depends on dimension. For instance at
$d=2$,
these are $qp_{c} (bond) \simeq 2.0$ and $fp_{c} (site) \simeq 0.45$, with $f$
the lattice filling factor. Extension to quasilattices [6, 7]
consists in replacing $q$ by
mean coordination number $\overline{z}$ in the invariant appropriate to bonds
(the problem for
sites is not solved, because there is no ambiguous definition of the filling
factor in quasicrystals [7]).

In the same spirit, we propose to extend Eq. (1), replacing $q$ by an
effective value
$q_{eff}$. The dependence of the percolation thresholds on site
connectivity does not
imply that the relevant variable should be the arithmetic average
$\overline{z}$.
Therefore, we regard $q_{eff}$ as a parameter which has to be determined,
rather
than arbitrarily forced to be equal to $\overline{z}$.

The need of substituting $q$ by $q_{eff}$ in the equations is easily
understood in both the case of quasilattices and the case of lattices which mix
different
values of $q$. It is also needed in anisotropic lattices. Let us again
consider the hexagonal lattice with spacing $a$, $b$ and $c$ in the three
lattice directions.
In the limit $c\rightarrow +\infty$, one is left
with decoupled $ab$ planes for physical systems having a finite
range of interaction. Therefore, the percolation
threshold of the hexagonal lattice must depend somehow on the ratio $c/a$.
It is expected to range between that of isotropic $bcc$
with $d=3,q=8$ and that of the
$d=2$ triangular lattice, associated to the limit $c/a \rightarrow +
\infty$. Note in this
limit one recovers an isotropic lattice with $q=6$ instead of $q=8$, which
suggests
an effective coordinance $6 \leq q_{eff} \leq 8$. In view of such
considerations,
we propose to extend Eq. (1) substituting $q$ by some $q_{eff}$ for any
lattice.

Note within the two classes defined by the set $(a,p_{0})$, $q_{eff}$ is
the unique
unknown parameter.
For each lattice, we find a value of this fitting parameter which, when
inserted
into Eq.1, reproduces both the site and the bond percolation threshold within
$\mp 0.01$. Values of $q_{eff}$ are reported for different lattices in
Table (1),
together with site and
bond
percolation thresholds $p_{c}^s$, $p_{c}^b$ obtained when $q$ is replaced
by $q_{eff}$
in Eq. (1). Mean coordination $\overline{z}$ and
$p_{c}^e$ for site and bond percolation are also reported for comparison.
Exact or
numerical estimates $p_c^e$ are from Refs [6, 7, 10].

%%%%%%%%%%%%%%%%%%%%%%%
\section{The results}

For
lattices at $d=2$, $q_{eff}$ differs from $\overline{z}$ by $1\%$
in the case of
the dual latttice of Penrose, and is smaller than $0.5\%$ for all the other
lattices.
These data corroborate that two dimensional lattices are divided into the
two distinct
classes defined in [1]. This is illustrated in Figs. (2) and (3), where
$p_{c}^e$'s are
reported in a $log-log$ plot such that the experimental points for
lattices in the same
class align on a straight line according to Eq.1.

The pertinent variable after Eq.
(1) is
$(d-1)(q_{eff}-1)$ for sites and $(d-1)(q_{eff}-1)/d$ for bonds. Those are
the variables
in abscissa in Fig. (3) which reports data for lattices in dimensions
$d=2$
and $d=3$ (data in higher dimensions have been already displayed in a
similar plot
in [1] with $q_{eff}=q$). Since the first class concerns only lattices
which are all at $d=2$, above variables for respectively sites and bonds
may reduce
to one single common variable
$q_{eff}$ as shown in Fig. (2).
 $\mid \Delta \mid$ reaches $0.01$
only in the
dodecagonal lattice with ferromagnetic links, and in the dual lattice of
Penrose. In all
the other cases, the error in the percolation threshold estimates is
only on the
third decimal.

Note the larger error in the dodecagonal lattice. We attribute it to an actual
bond percolation threshold $p_{c}^{e}(bond) = 0.495$ [6] larger than a
priori expected.
One would indeed
have expected $p_{c}(bond) = 0.475$ from the straight line in Figure 3, in
which case a value
$q_{eff}=4.289$, close to $\overline z = 4.27$, would have reproduced both
site and
bond percolation
thresholds within $\mid \Delta \mid = 0.001$.
The problem with the estimate $p_{c}^{e}(bond) = 0.495$ is also evident
from the Scher and
Zallen invariant. It yields
$\overline z p_{c}^{e}(bond) = 2.11$, the largest value of this parameter
among
all the lattices investigated [6]. At the opposite,
$p_{c}(bond) = 0.475$ would yield $\overline z p_{c}(bond)
= 2.028$, close to the invariant value 2.0 at $d=2$.

At $d=3$, percolation values are reproduced with a very good
accuracy, since
$\mid \Delta \mid \leq 0.004$ for all the lattices. The difference between
$q_{eff}$ and
$\overline z$ is not negligible as it reaches few per cent in some cases.
This is easily
understood in the case of the hexagonal (stack triangle) case, where we
have argued
earlier that a value $q_{eff}$ smaller than $8$ is expected. Actually, we find
$q_{eff}=7.66$, as a consequence of the anisotropy. The hexagonal close packed
($hcp$) lattice has
percolation thresholds which are different from those of the $fcc$ lattice,
as
expected since the $hcp$ lattice is not a Bravais lattice.

However, differences are
small which may be attributed to the fact that both lattices are indeed
isotropic, each
site being surrounded by $12$ equivalent neighbors. In this context, the
small value of
$q_{eff}$ close to $11$ in $hcp$ lattice is not only due to the  non-Bravais
nature of the
lattice. It is also related to the fact that in the $fcc$ lattice, $q_{eff}$
is only $11.6$,
significantly smaller than  $q=12$, although the coordination number is
single-valued.

On the other hand, $q_{eff}$ differs from $q$ by only $1\%$ in the other
lattices ($sc$, diamond,
$bcc$). Yet one would expect $q_{eff}=q$ in such isotropic lattices
with single-valued coordination number.
The difference
between $q_{eff}$ and $q$ in this case illustrates that
our formula for the
percolation thresholds is not exact as we
already stated in [1], and shown convicingly in ref. 2. Nevertheless,
both site and bond percolation thresholds for all the
lattices in any dimension are provided within $1\%$ by our universal law
involving only two parameters: the dimension $d$ and a parameter $q_{eff}$
which contains more information on the geometry than the mean coordination.

%%%%%%%%%%%%%%%%%%%%
\section{Conclusion}

We  have shown that our universal formula for percolation threholds we
reported
earlier [1] for periodic Bravais lattices with equivalent nearest neighbors
does extend
to any kind of lattice, provided the coordination number is replaced by an
effective value $q_{eff}$. We then conclude that a good estimate of
both site and bond percolation thresholds can be obtained from the formula
in Eq. 1 involving only the dimension $d$ of the lattice, and one parameter
$q_{eff}$ which contains the geometric information of the lattice. This
parameter however, does not reduce to the mean coordination number
$\overline z$,
although $\overline z$ and $q_{eff}$ differ by few per cent only.
Indeed,
we find the universality does include
different numerical
values for $q_{eff}$ and thus different percolation
thresholds for lattices which have the same set ($d,\ \overline z$). This is
evident from Table (1)
which reports results for as many as six lattices with $d=2,\overline z
=4$, two lattices
with ($d=3,\ \overline{z}=8$), two other ones with ($d=3,\ \overline{z}=12$).

In our previous work, we had only three lattices belonging to the first class.
The present work extends this class from 3 to 8 lattices. We might have
invoked
chance for three
points aligned on a same line, but not for
8 points like in Fig.1. Therefore, we confirm
the existence of two different classes, one for some of
the two-dimensional
lattices, the other one for all the other two-dimensional
lattices and all the lattices up to d=7.

We do not have a scheme
to derive the relevant variable $q_{eff}$ from the
geometry of the lattice. However, an important result of this work is that
this variable does exist. It  means that there exists a single value for
$q_{eff}$ which accounts for
both the site and the bond percolation thresholds for any given lattice within
$\mp 0.01$.
This result is sufficient to give our formula a prediction ability for
pecolation thresholds of other lattices which have not been computed yet.
For example, the knowledge of one site (bond)
percolation  threshold for a given lattice is sufficient to determine
a point on the relevant straight line in Fig. 2 or 3. Then
$q_{eff}$ can be found
from the absissa of this point, which in turn allows for the determination
of the bond (site) percolation threshold from our universal formula, within
$\mp 0.01$. Depending on whether $p_c^s$ or $p_c^b$ is known, the value of
$q_{eff}$ deduced from Fig. (2) or (3) will be different. However, this
difference only corresponds to the deviation of the $p_c^e$'s with repect
to the universal law, i.e. less than 1\% with the exception of only
few out-liers.

The robustness of our formula suggests the extension
to more complex problems such as
directed percolation. Also, anisotropic percolations with different bond
probabilities in different directions may be adressed in the near future.

%%%%%%%%%%%%%%%%%%%%%%%%%%%

\subsection*{Acknowledgments.}
We would like to thank Dietrich Stauffer for
stimulating comments.

%%%%%%%%%%%%%%%%%%%%%%%%%%%%%%%%%%%%%
\newpage
\vspace{3.0cm}
{\LARGE References}\\ \\
1. {\sf S. Galam and A. Mauger}, Phys. Rev. E \underline {53}, 2177 (1996). \\
2. {\sf Van der Marck}, Phys. Rev. E \underline {55}, 1228 (1997);
{\sf S. Galam and A. Mauger}, Phys. Rev. E \underline {55}, 1230 (1997). \\
3. {\sf V.K.S. Shante and S. Kirpatrick}, Adv. Phys. \underline {20}, 326
(1971). \\
4. {\sf J. P. Lu and J. L. Birman}, J. Stat. Phys. \underline{40}, 1057
(1987).\\
5. {\sf F. Babalievski and O. Pesshev}, C. R. Acad. Sci. Bulg. \underline{41},
85 (1988).\\
6. {\sf F. Yonesava, S. Sakamoto, K. Aoki, S. Nose and M. Hori}, J.
Non-Cryst. Solids
\underline{123}, 73 (1988).\\
7. {\sf F. Babalievski}, Physica \underline{220}, 245 (1995).\\
8. {\sf D. Leduie and J. Teillet}, J. Non-Cryst. Solids \underline{191},
216 (1995).\\
9. {\sf H. Scher and R. Zallen}, J. Chem. Phys. \underline{53}, 3759 (1970).\\
10. {\sf D. Stauffer and A. Aharony}, {\em Introduction to Percolation
Theory}, 2nd ed.
(Taylor and Francis, London, 1994).

\newpage
{\LARGE Figure captions}\\ \\

Fig. 1. Less common lattices studied in this paper : Kagom$\acute{e}$ (a)
and its dual,
Dice lattice (b); Penrose quasicrystalline lattice (c) and its dual (d);
octagonal
quasicrystalline lattice with ferromagnetic links (e) and chemical links
(f); dodecagonal
quasicrystalline lattice with ferromagnetic links (g) and chemical links (h).

Fig. 2. Inverse of percolation thresholds as a function of
the variables
$(d-1)(q_{eff}-1)$
and $(d-1)(q_{eff}-1)/d)$ which reduce here ($d=2$) to the single variable
 $q_{eff}$ in logarithmic scales for lattices belonging to the first
class.

Fig. 3. Inverse of percolation thresholds as a function of the variables
$(d-1)(q_{eff}-1)$
and $(d-1)(q_{eff}-1)/d)$ approriate to site and bonds, respectively, in
logarthmic scales for lattices belonging to the second class.

%%%%%%%%%%%%%%%%%%%%%%%%%%%%%
%%%%%%%%%%%%%%%%%%%%%%%%%%%%%
\newpage
\def\figure2{\picture 170.mm by 165.mm (firstclass)}
\centerline{\figure2}
\newpage
\def\figure3{\picture 170.mm by 165.mm (seconclass)}
\centerline{\figure3}

%%%%%%%%%%%%%%%%%%%%%%%%%%%%%%%%%%%%%

%%%%%%%%%%%%%%%%%%%%%%%%%%%%%%%%%%%%
%%%%%%%%%%%%%%%%%%%%%%%%%%%%%%%%%%%%%%%%%%%%%%%%%%%%%%%%%%%%%%%%%%%%%%%
\newpage
\begin{table}

\label{tbl}
\begin{center}
\begin{tabular}{|c|c|c|c|c|c|c|c|c|}
\hline
\multicolumn{9}{|c|}{first class}\\
\hline
& &
&\multicolumn{3}{|c|}{site}&\multicolumn{3}{c|}{bond}\\ \cline{4-9}
\multicolumn{1}{|c|}{Lattice} &\multicolumn{1}{|c|}{$\overline z$}
&\multicolumn{1}{|c|}{$q_{eff}$}
&\multicolumn{1}{|c|}{$p_c^e$}
&\multicolumn{1}{|c|}{$p_c$}
&\multicolumn{1}{|c|}{$\Delta$}
&\multicolumn{1}{|c|}{$p_c^e$}
&\multicolumn{1}{|c|}{$p_c$}
&\multicolumn{1}{c|}{$\Delta$} \\
\hline
Square &4& 4.02&  0.5928& 0.5970 & +0.0042 & 0.5 & 0.4935 & -0.0064\\
Honeycomb & 3&  2.99& 0.6962  & 0.6938 & -0.0024 & 0.6527 & 0.6581 & +0.0054\\
Triangular & 6& 5.98 & 0.5 & 0.4986 & -0.0014 & 0.34729 & 0.34955 & +0.0023\\
Dice$\ast$ & 4 & 4.189 & 0.5851 & 0.5854 & +0.0003 & 0.476 & 0.4754 & -0.0006\\
Penrose$\ast$ & 4 & 4.194 & 0.5837 & 0.5851 & +0.0014 & 0.477 & 0.4748 &
-0.0022\\
Octa.chem.links$\ast$ & 4 & 4.170 & 0.585 & 0.5867 &+0.0017 &0.48 &0.4773 &
-0.0027\\
Octa.ferro.links$\ast$ & 5.17 & 5.013 & 0.543 & 0.5389 & -0.0041 & 0.40 &
0.406 & +0.0057\\
Dode.chem.links$\ast$ & 3.63 & 3.638 & 0.628 & 0.6269 & -0.0011 & 0.54 &
0.5419 & +0.0019\\
\hline
\hline
\multicolumn{9}{|c|}{second class}\\
\hline
& &
&\multicolumn{3}{|c|}{site}&\multicolumn{3}{c|}{bond}\\ \cline{4-9}
\multicolumn{1}{|c|}{Lattice} &\multicolumn{1}{|c|}{$\overline z$}
&\multicolumn{1}{|c|}{$q_{eff}$}
&\multicolumn{1}{|c|}{$p_c^e$}
&\multicolumn{1}{|c|}{$p_c$}
&\multicolumn{1}{|c|}{$\Delta$}
&\multicolumn{1}{|c|}{$p_c^e$}
&\multicolumn{1}{|c|}{$p_c$}
&\multicolumn{1}{c|}{$\Delta$} \\
\hline
Kagom$\acute{e}$&  4& 3.980 & 0.6527 & 0.6567 & +0.0040 & 0.5244 & 0.5195 &
-0.0049\\
dual Penrose& 4 & 4.04 & 0.6381 & 0.6487 & +0.0106 & 0.5233 & 0.5099 &
-0.01341\\
Dode.ferro.links$\ast$ & 4.27 & 4.218 & 0.617 & 0.6264 &+0.0094 & 0.495 &
0.4835 & - 0.0115 \\
hex. compact & 12 & 11.146 & 0.204 & 0.2015 & -0.0025 & 0.124 & 0.1263 &
+0.0023\\
stac. triangle & 8 & 7.661 & 0.2623 & 0.2611 & -0.0012 & 0.1859 & 0.1872 &
0.0013\\
Diamond & 4& 4.0087 & 0.43 & 0.4260 & -0.0040 & 0.1859 & 0.3935 & +0.0055\\
sc & 6 & 5.9558 & 0.3116 & 0.3132 & +0.0016 & 0.2488 & 0.2468 & -0.0020\\
bcc & 8 & 8.1355 & 0.246 & 0.2502 & +0.0042 & 0.1803 & 0.1755 & -0.0047\\
fcc & 12  & 11.626 & 0.198 & 0.1958 & -0.0022 & 0.119 & 0.1210 & 0.0020\\
\hline
\end{tabular}
\end{center}
\caption{\sf percolation thresholds from this work $p_c$ compared to
``exact estimates"
$p_c^e$ taken from [6, 7 and 10]. $\Delta \equiv p_c-p_c^e$. $\ast$ refers to
multi-valued coordination number (mean coordination $\overline z$).
 All the lattices are in dimension $d=2$, except the 6 last ones in the
second class
which have $d=3$.
We have completed the table caption as follows:
$q_{eff}$ has been chosen as the arithmetic average of the parameters $q$
 which reproduces $p_c^e$ for respectively site and bond when
inserted in Eq. (1).}
\end{table}

\end{document}